# A novel approach to combine spatial and spectral information from hyperspectral images.

**Belal Gaci [1,2,3], Florent Abdelghafour [2,3], Maxime Ryckewaert[2,3], Silvia Mas-Garcia [2,3], Marine Louargant [1], Florence Verpont [1], Yohana Laloum [1], Ryad Bendoula [2,3], Gilles Chaix[3,4,5], Jean-Michel Roger [2,3].**

[1] CTIFL, France
2 ITAP-INRAE, Institut Agro, University Montpellier, Montpellier, France
[3] ChemHouse Research Group, Montpellier, France
[4]CIRAD UMR AGAP

**Abstract**

# 1 Introduction

Hyperspectral imaging is an image acquisition technique that allows for collecting data over a wide range of electromagnetic wavelengths (Khan et al., 2018). Hyperspectral images contain information about the different types of materials present in the imaged scene, which can be used for object classification (Folch-Fortuny et al., 2016), vegetation mapping (Lelong et al., 1998), environmental monitoring, and other applications. This technique is a prolific research topic (Feng and Sun, 2012). A hyperspectral image (HSI) is defined as a hypercube consisting of images acquired of the same scene but at different wavelengths. Based on this definition, an HSI can be considered as a hypercube organized into three modes. The first two modes are related to the image geometry (x, y), and the third mode is related to the wavelengths (λ) of the spectra (Imani and Ghassemian, 2020). HSI provide rich information but are complex to process due to the high dimensionality of the spaces they belong to. Different methods have been proposed to solve this complexity. Most of them focus on the spectral dimension and often neglect the spatial dimension. In these methods, the images are unfolded as a matrix of spectral samples and processed using chemometric tools (Amigo et al., 2015). With this process, the spatial information contained in the HSI, i.e., the arrangement of spectra to form patterns, is not considered. However, in many cases, objects in an HSI have very similar spectral signatures but different shapes and textures (Ghassemian Yazdi, 1988). In these situations, it is important to add spatial information to differentiate these objects. Methods have recently been developed to consider the local properties of the image. They aim to integrate spatial features into the processing to improve the efficiency and reliability of the models built with HSI (Imani and Ghassemian, 2020). This can be done in two approaches. The first approach is to separately extract spatial and spectral features and then fuse them with a combination method (Imani et al., 2017). Among these methods, Xu (2020) proposes to reduce the HSI with a principal component analysis, then to calculate spatial features on the resulting images, and to combine them with spectra (Xu, 2020). Another method is to associate a local rank matrix with pure spectra to determine the components that exist in an HSI (De Juan, 2019). The second approach is to jointly extract spatial and spectral features. Some of these methods introduce spatial constraints into the algorithm analysing the HSI in the spectral domain, particularly multivariate curve resolution methods (Vitale et al., 2020). More recently, some deep learning methods, initially dedicated only to spatial features, have been extended to HSI processing (Paoletti et al., 2019). All these methods are based on modifying or adapting purely spectral or purely spatial methods to the case of HSI.

The approach proposed in this article aims to solve the question of jointly processing the spatial and spectral information of an HSI symmetrically, without relying on an existing method. It proposes a methodological framework consisting of estimating spatial features with monochromatic imaging tools and spectral signatures with chemometric tools, and then fusing them.

The first part presents the theoretical framework of this approach. The second part presents the materials and methods of two case studies used to illustrate this method. The third part presents and discusses the results obtained.

## 2 Theory

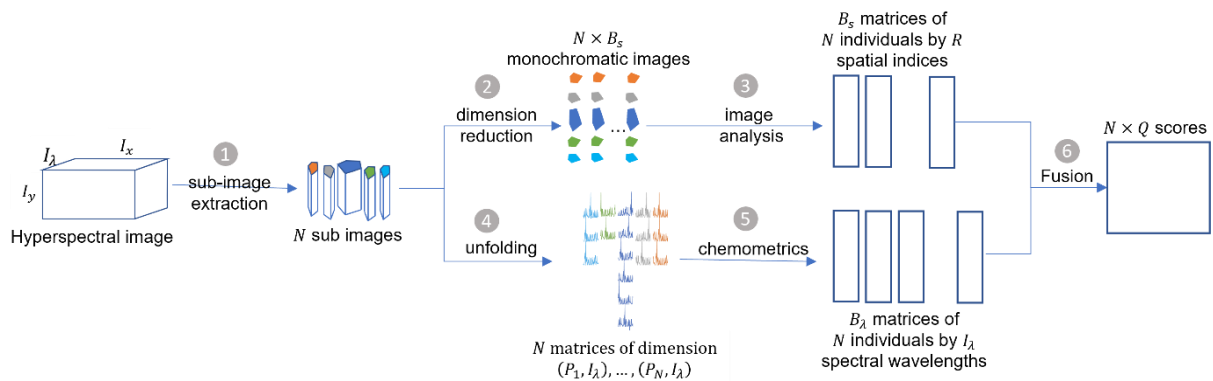

*Figure 1: General scheme of the spatial-spectral method of processing hyperspectral images*

The method illustrated in Figure 1 proposes to process a hyperspectral image of width $I_x$, height $I_y$, acquired on $I_\lambda$ wavelengths, by extracting spatial and spectral information at the scale of N regions of interest or sub-images (step 1). The aim of this method is to characterize each sub-image by a set of scores. Different types of sub-images can be considered. If the image is composed of objects, each sub-image can correspond to an object. Otherwise, the sub-images can be arbitrary geometric shapes. In this article, the sub-images are odd-width squares, in order to characterize the central pixel.

To extract the spatial and spectral information, two processing paths are used. The first path (upper branch of Figure 1) focuses on the spatial dimension. To do this, the spectral dimension is reduced to a small number of $B_s$ variables (step 2). Different approaches of dimension reduction are possible. In this paper, principal component analysis, PCA (Wold et al., 1987) was used. This method defines a set of factors corresponding to linear combinations of the wavelengths. These factors are determined by performing PCA on a training set, which can be composed of all the spectra of the image to be processed or any other external training set. The projection of the spectra of each sub-image onto these factors allows them to be reduced to a set of scores, thus producing a reduced number of monochromatic sub-images. Then, the $N \times B_s$ monochromatic sub-images are analysed to extract representative spatial features (step 3). This operation can be based on various image analysis techniques. Since the method presented in this article aims to characterize the central pixels of the sub-images, texture analysis was chosen, using structure tensors (Bigun et al., 2004) for the first case study and gray-level co-occurrence matrices (GLCM) combined with Haralick descriptors (Haralick, 1979) for the second case study. Both methods produce $R$ indices characterising the texture of each sub-image. At the end of this step 3, $B_s$ matrices of $N$ individuals described by $R$ variables are obtained.

The second path (lower branch of Figure 1) focuses on the spectral dimension. To do this, the $N$ sub-images are unfolded to obtain $N$ matrices (step 4). The number of rows of each matrix is equal to the number of pixels of each sub-image (P_1… P_N) and the number of columns is equal to the number of wavelengths $I_\lambda$. Then, the $N$ matrices are analysed to extract representative spectral signatures of the spectral variations carried by each patch (step 5). This operation can be based on various statistical and chemometric techniques, which produce spectral signatures or loadings. In this article, the average of the patch spectra was chosen for the first case study. For the second case study, a singular value decomposition (SVD) (Klema and Laub, 1980) was performed on each matrix, without centring, and the first loadings were retained.

The spatial and spectral information calculated by the two branches of Figure 1 are presented in the form of $(B_s + B_\lambda)$ matrices of N rows, which are then merged (step 6). The fusion of low and medium level data consists of modelling from the data sources, while the fusion of high-level data combines the responses of models built for each data set (Doeswijk et al., 2011). Low-level fusion is proposed here, as it requires few parameter settings. Different low-level fusion methods have recently been developed, under the name of multi-block methods. The choice of the multi-block method depends on several factors. First, if the model being sought is exploratory (unsupervised) or predictive (supervised). Second, whether the number of blocks to be merged is large or not, which depends on the parameters $B_s$ set $B_\lambda$. In this paper, an unsupervised fusion is performed using MB-PCA (Hanafi et al., 2020) for the first study case and a supervised fusion is performed using PLS-ROSA (Liland et al., 2016) for the second study case. This second method has the advantage of being able to handle many blocks.

## 3 Materials et Methods

### 3.1 First case study: wood disk

The study used a sample from a plantation owned by the Brazilian company Floresteca in the state of Mato Grosso, Brazil. The samples were prepared at the University of São Paulo (ESALQ) in Piracicaba, São Paulo state, Brazil. A 3 cm thick transverse disk was taken from a teak tree at chest height. To obtain a perfectly smooth surface, one side of the disk was sanded with 120-grit sandpaper. The disk was then placed in a climatic chamber (20°C, 60% RH) to reduce and homogenize its moisture content to 12-15% within 48 hours at 20°C and 60% relative humidity.

#### 3.1.1 Image acquisition and pre-processing

Hyperspectral images in the near infrared (NIR-HSI) were obtained using a stationary chemical imaging camera (SisuCHEMA, SWIR, Specim®) with a Specim OLE15 f/2.0 lens (field of view of 200 mm) and a linear scanning system of 63.3 mm/s, operating in a wavelength range of 928 to 2524 nm with 6 nm intervals. This wavelength range was segmented into 256 spectral channels, resulting in a hyperspectral image cube of 256 wavelengths. Images were acquired on the polished transverse surface of the disks with a pixel size of 625 x 625 µm. The reflectance of the sample was calculated using a white reference standard and a dark internal reference, and the data was then transformed into absorbance using a logarithmic function.

### 3.1.2 Data analysis

A second order derivative was applied to all spectra in the image. The Savitzky Golay algorithm (Savitzky and Golay, 2002) was used with a window of width 13 and a polynomial of degree 3. The 15 wavelengths at the two ends of the spectra, which were affected by the edge effects of the derivation, i.e., from 964 to 1072 nm and from 2428 to 2494 nm, were removed, reducing the number of wavelengths used to $I_\lambda = 226$. A dataset of N = 54290 sub-images of size 7x7 was formed. The central pixels of these sub-images were introduced into a principal component analysis. The first loading was placed in a vector **p** $(226 \times 1)$.

The methodology described in section 2 was applied to this dataset. For the spatial analysis (upper branch of Figure 1), each hyperspectral sub-image was projected onto the vector **p**, producing N monochromatic sub-image (step 2, $B_s = 1$). Then, a local structure tensor was calculated on each sub-image. From this tensor, variables of magnitude, phase, and coherence were calculated (Medioni, 2000) (step 3). At the end of this operation, a spatial block of dimension $(N = 54290 \times R = 3)$ was obtained.

For the spectral analysis, each sub-image was unfolded into a matrix of 49 spectra (step 4). The mean spectrum of each of these matrices was calculated (step 5). At the end of this operation, a spectral block of dimension $(N = 54290 \times I_\lambda = 226)$ was obtained. Then each block was normalized with respect to its global standard deviation and centred before being introduced into an unsupervised analysis of the MBPCA type (M Hanafi et al., 2020) (step 6). The first 7 components were retained, producing 7 score images.

### 3.2 Second case study: apple scab

The reader can find more details about the plant material, inoculation, and experimental setup in (Nouri et al., 2018) and (Gorretta et al., 2019).

### 3.2.1 Plant material

The study was conducted on two potted apple trees (X months old, "rootstock X", cv Golden Delicous). The selected cultivar was Golden Delicious, which is considered moderately susceptible to apple scab disease caused by *Venturia inaequalis*. One tree was infected and the other was non-inoculated (control). The infected plants were inoculated by spraying a suspension of $5.10^5$ conidiospores per milliliter using a commercial manual sprayer. The suspension was obtained by collecting the spores from apple scab infected leaves. The control plant were kept healthy and sprayed with water. The plants were then placed under optimal incubation conditions with 100% relative humidity, at 18°C, and growing in darkness during 48 hours to stimulate fungal development.. The plant was evaluated daily by an expert who visually noted the development of apple scab symptoms. On the last day of the experiment, the inoculated leaf was detached from the plant and examined by the expert using a magnifying glass to identify the positions of the apple scab brown spots.

### 3.2.2 Image acquisition and pre-processing

The images were acquired with a HySpex SWIR-320m-e hyperspectral camera (Norsk Elektro Optikk, Norway). The images were acquired line by line with a step of 0.287 mm. Each line consists of 320 pixels. Each pixel is composed of a spectrum of 256 wavelengths ranging from 960 nm to 2490 nm with a step of 6 nm. On days 2, 3, 4, 5, 6, 9, 10, and 11 after inoculation, images of the inoculated leaf and the control leaf were acquired under laboratory conditions. The leaves were fixed on an appropriate support to avoid damaging the plant material. A halogen source illuminated the leaves homogeneously. The images were transformed into reflectance (Kortüm, 2012) using a white reference placed in each image.

Eight symptomatic zones were identified and labelled on the images of the inoculated leaf (see Nouri et al, 2018). Similarly, eight control zones were labelled on the images of the control leaf. These zones correspond to parts of the blade outside the main veins, of equal size and geometry to those of the symptomatic zones. The positions of the control and symptomatic zones were the same on each date. For each zone, all square images of size 3x3 included in the zone were extracted (cf. Table 1).

*Table 1 : Number of sub-images per zone, for each date.*

| Zone number | Number of sub-images for the inoculated leaf | Number of sub-images for the control leaf |
|---|---|---|
| 1 | 28 | 28 |
| 2 | 144 | 144 |
| 3 | 19 | 19 |
| 4 | 5 | 5 |
| 5 | 5 | 5 |
| 6 | 1 | 1 |
| 7 | 2 | 2 |
| 8 | 4 | 4 |
| TOTAL | 208 | 208 |

### 3.2.3 Data analysis

A calibration set was created with days 2, 4, 6, 9, and 10, and a test set with days 3, 5, and 11. The calibration set contained N = 2080 sub-images, including 208 sub-images of the uninoculated leaf and 208 sub-images of the inoculated leaf repeated over 5 days. The spectra of the central pixels of the sub-images in the calibration set were input into a principal component analysis. The first $B_s = 3$ principal components, explaining 99% of the variance, were retained and placed in a matrix **P**.

The methodology described in Section 2 was applied to the calibration and test sets. For the spatial analysis (upper branch of Figure 1), each hyperspectral sub-image was projected onto the first 3 loadings of matrix **P**, producing $N \times 3$ monochromatic sub-images (Step 2, B_s = 3). These sub-images were then transformed into GLCM matrices. Three parameters were considered for the GLCM calculation: the distance $d$ between two pixels, the orientations $\theta$ between pixel pairs, and the number of gray-level classes. The distance was fixed at $d = 1$. Four orientations were considered for $\theta = [0°, 45°, 90°, 135°]$. Then, the four matrices were added to obtain an isotropic matrix. For the number of gray-level classes, values between 2 and 19 were tested. The $R = 14$ Haralick indices were then calculated for each sub-image (Step 3). At the end of this operation, 3 spatial blocks of dimension ($N = 2080 \times R = 14$) were obtained.

For the spectral analysis, each hyperspectral sub-image was unfolded into a matrix of 9 spectra (Step 4). An SVD was performed without centring the spectra matrix, and the first $B_\lambda = 3$ loadings were retained as spectral features (Step 5). At the end of this operation, $B_\lambda = 3$ spatial blocks of dimension ($N = 2080 \times I_\lambda = 256$) were obtained.

These blocks were then used to construct a multi-block regression model, ROSA-PLS (Liland et al., 2016) (Step 6). The response to be predicted was 0 for pixels in control areas and 1 for pixels in symptomatic areas. The scores produced by the ROSA model were then input into a LDA classification method (Brown and Tinsley, 1983) to produce the probabilities of membership in the two classes, "control" and "infected." The performance of the models was characterized based on a confusion matrix.

### 3.2.4 Calibration and test procedures

The estimation of optimal parameters for ROSA-PLS was determined by a double k-fold cross-validation (Filzmoser et al., 2009) applied within the calibration set. The k-fold cross-validation involves dividing the sample into k groups. One of the groups is then used as a validation sample and the remaining k-1 groups represent the training set. This operation is repeated k times, changing the validation sample each time, to test all groups. The double cross-validation was defined as two nested loops. For the inner loop, each group was composed of samples from the same infected spot and independent control region for a given date. The inner loop was composed of 40 groups. For the outer loop, each group was composed of 20 groups randomly selected from the 40 groups in the inner loop. The selected model corresponded to the number of latent variables that minimized the classification error by double cross-validation after the LDA decision. Then, this model parameterized with the optimal number of latent variables was applied to the test set.

## 4 Results and discussion

### 4.1 First case study:

This first case study is an application of the unsupervised method for the fusion of spatial and spectral variables, to teak wood disks hyperspectral images. Figure 2 shows the 7 score images and their corresponding loadings obtained with the MBPCA method.

Image score 1

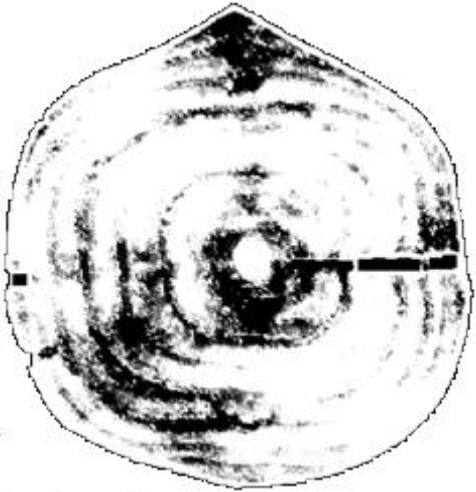 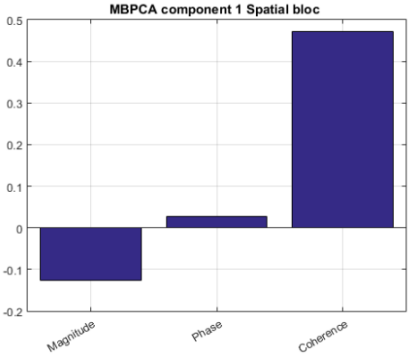 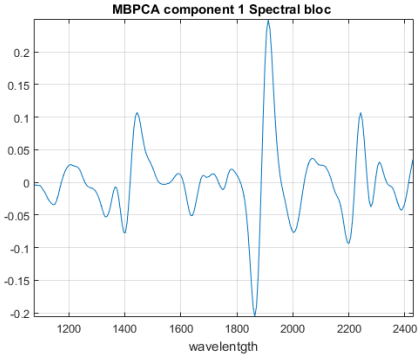

Image score 2

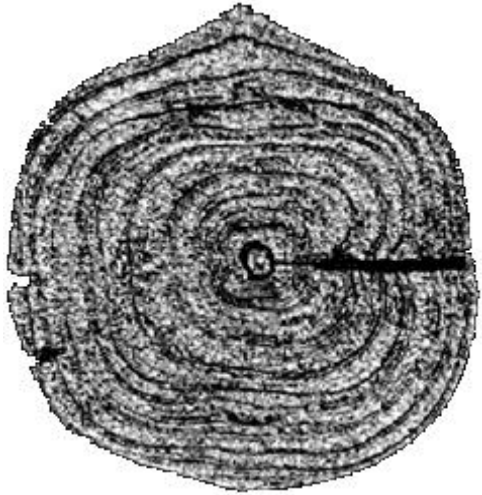 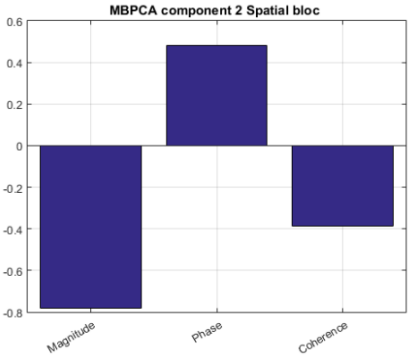 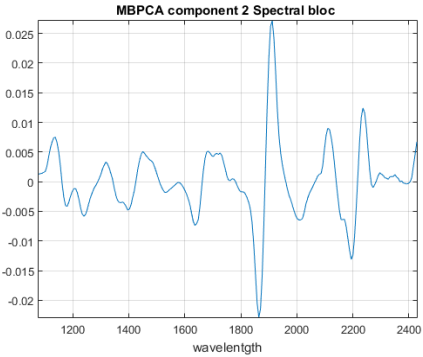

Image score 3

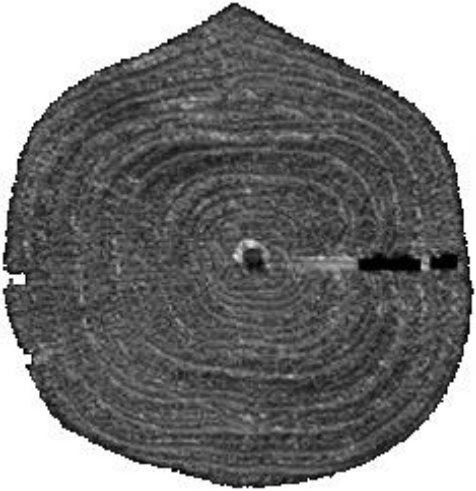
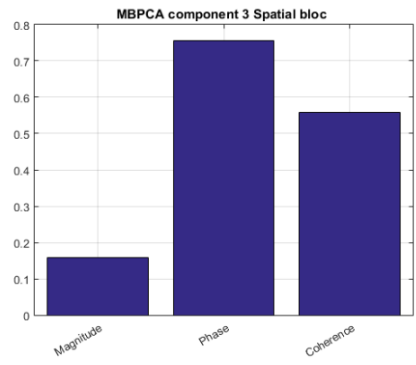
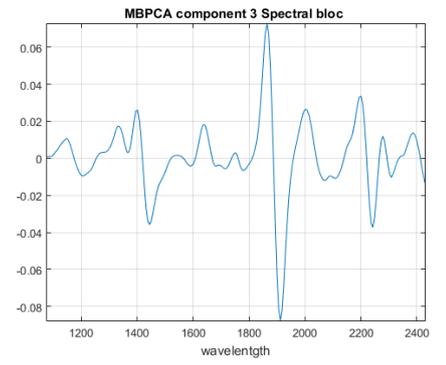

Image score 4

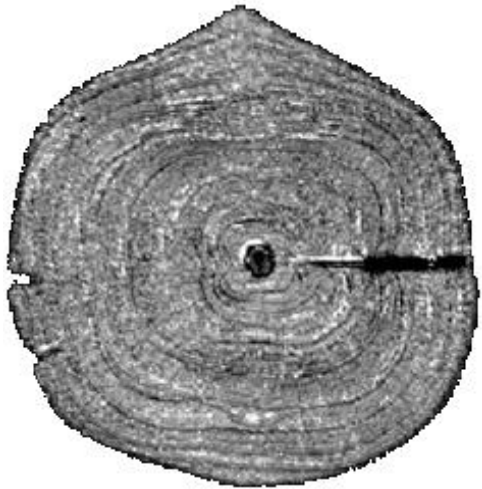
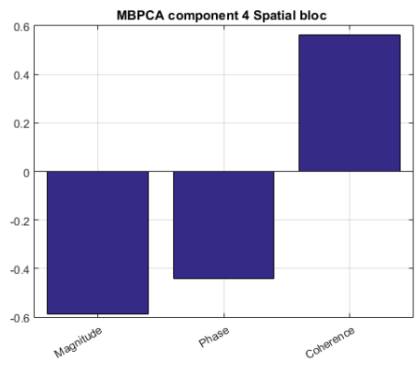
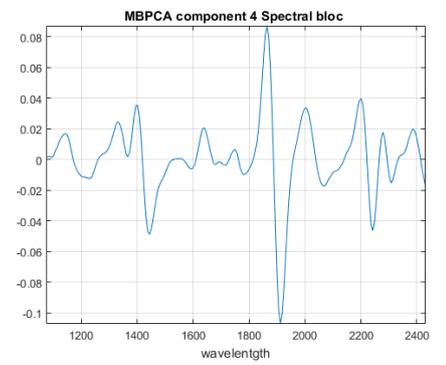

Image score 5

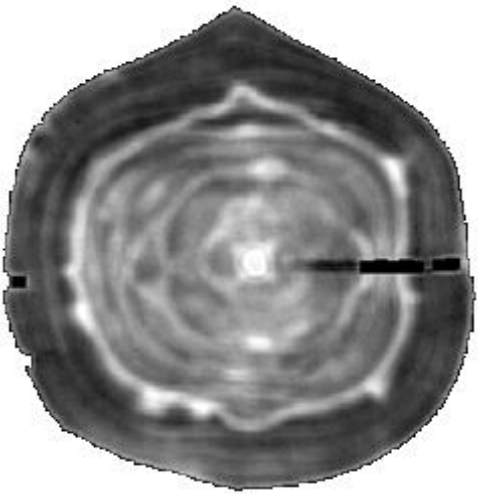
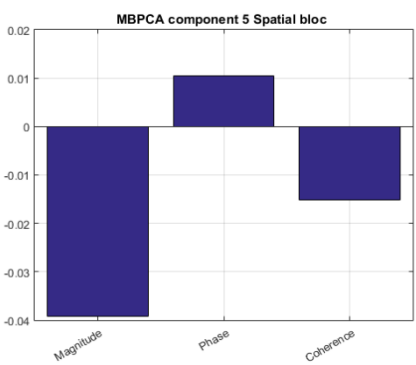
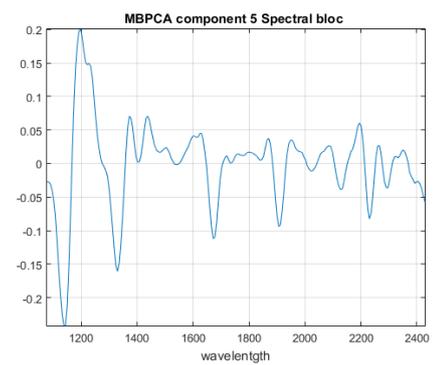

Image score 6

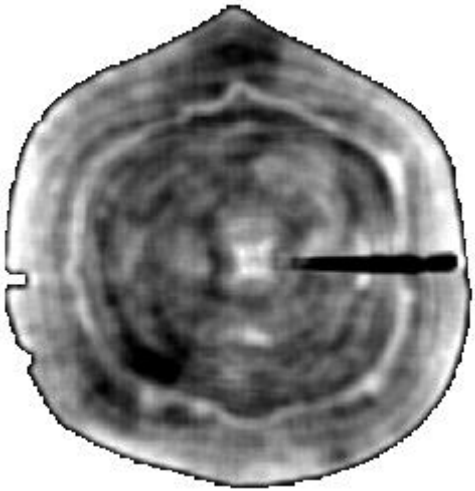
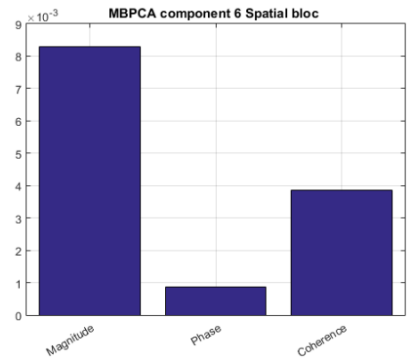
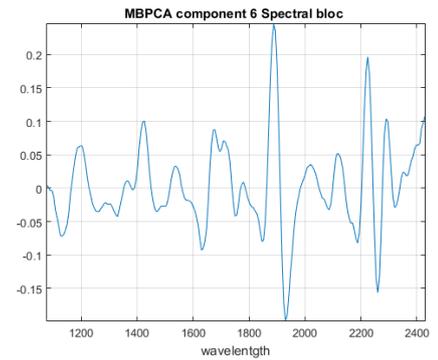

Image score 7

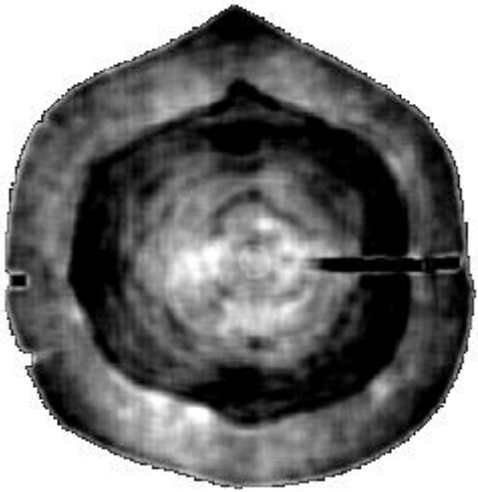
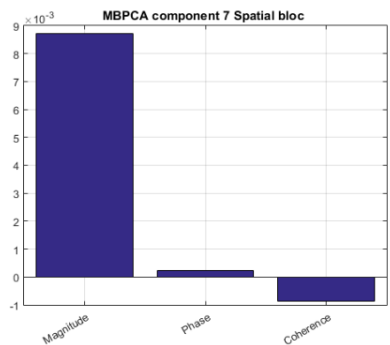
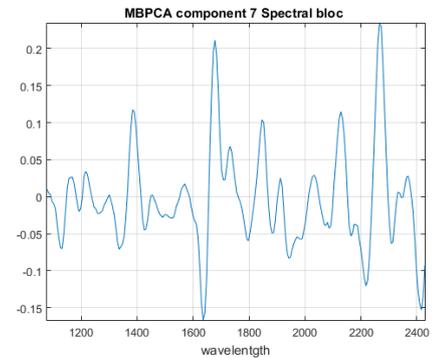

*Figure 2: On the left, MBPCA score images. In the middle the spatial loadings. On the right the spectral loadings*

The different score images and loadings are interpreted in the following paragraphs. For score images 1, 3, and 4, the spectral loadings resemble the mean spectrum (Figure 3(a)) of the images (with the sign reversed for image 1). Therefore, they reflect a global variation in the luminance within the hyperspectral images. This is certainly due to a heterogeneity of the overall reflective properties of the wood. From an optical point of view, these variations may be related to variations in particle size or optical indices of the cell walls (Griffiths and Dahm, 2007).

Image score 1:

This image presents two dark areas. The first is located at the top of the image, and the second runs from the pith towards the left end of the wood disk. From a spatial point of view, the coherence variable shows a strong weighting. This variable expresses the anisotropy, i.e., the dark areas exhibit strong local anisotropy in terms of overall luminance, while the light areas are more isotropic. The dark area at the top could correspond to a branch departure, while the dark area at the bottom could correspond to tension wood. These areas may contain more filled and anisotropic wood than the rest of the disk.

Image score 2:

The second score image displays concentric circles with unclear contours, and with a variation of grayscale levels inside each circle, producing a gradient effect that darkens towards the boundary. There is also a darker ring surrounding the centre. Regarding the spectral part, several peaks are observed, including a positive peak at 1138 nm, which may correspond to the aromatic C-H absorption of lignin (Ma et al., 2019). A positive peak appears at 1210 nm corresponding to vibrations in the methyl (CH3) and methylene (CH2) groups of cellulose (Schwanninger et al., 2011). A positive peak appears at 1324 nm, which may correspond to the OH harmonic band of cellulose (Pan et al., 2021) (Via et al., 2003). A positive peak at 1912 nm may correspond to a lateral shift of the water peak. In addition, a negative peak is observed at 2200 nm, which corresponds to lignin. Finally, an unidentified negative peak appears at 1864 nm. This spectral profile thus highlights variations in cellulose and lignin. As for the spatial part, the three variables coherence, phase, and magnitude are significant, with magnitude having a stronger weight. The coherence and magnitude variables are negative, while the phase variable is positive. This spatial information, combined with the interpretation of the spectral peaks, shows that the dark areas, such as the boundaries of growth rings and the ring around the pith, are more homogeneous and isotropic, and vary in privileged directions in terms of cellulose. The light areas, such as the inside of the growth rings, are more heterogeneous and anisotropic in terms of lignin. These observations may be related to the variation in wood density, which is less dense at the beginning of the growing season than at the end.

Image score 3:

The third score image displays two overlapping patterns. The first, macroscopic pattern consists of concentric circles. The second, at a finer scale, consists of a succession of light, grey and dark points. In the centre, the pith shows lower scores than the rest of the image. In terms of spatial analysis, phase and coherence have significant and positive weights. This means that the bright areas located on the growth rings present more anisotropy in terms of overall luminance along preferred directions. In the bright areas, grains with strong local variation in phase and coherence are also observed. This represents a set of microstructures, either with gradients in all directions or with very directional gradients whose angle varies depending on their position in the image. Therefore, this image represents the variation in the number of conducting vessels and consequently the size of the fibres. Indeed, the size of the vessels is larger at the beginning than at the end of the season.

Image score 4:

The fourth image shows dark circles with white discontinuities, as well as a darker ring around the pith that appears as a dark grey colour. The inside of the growth rings displays a variation of lighter greys, with some white points near the boundaries. In terms of spatial features, all three variables have significant weights. However, magnitude and phase are negative, while coherence is positive. This indicates that the bright areas on certain parts of the growth rings exhibit homogeneous but anisotropic variations, while the dark areas, which make up most of the growth rings and the ring around the pith, are more heterogeneous and show variations in preferred directions in terms of global luminance. The dark areas correspond to the boundaries of annual growth rings, and the white spots may be associated with blocked vessels and fibres due to sanding of the wood disk (Figure 3(b) and (c)).

Image score 5:

The image shows two distinct areas, a dark area corresponding to the sapwood and a light area corresponding to the heartwood. The sapwood area is more homogeneous than the heartwood area, which displays discontinuous circular patterns. The spectral loadings present a negative peak at 1144 nm, corresponding to the lignin absorption band, and a positive peak at 1200 nm, corresponding to the starch absorption band (Williams and Norris, 1987), as well as a positive peak at 1444 nm, which also corresponds to starch absorption (Williams and Norris, 1987). A negative peak appears at 1678 nm, corresponding to polyphenols. However, an unidentified negative peak appears at 1330 nm, and the negative peak observed at 1906 nm do not correspond to any known compound in the literature. This last peak may correspond to a water peak trough that appears between 1916 nm and 1942 nm (Schwanninger et al., 2011). A negative peak is also observed at 2128 nm, corresponding to lignin, and a negative peak at 2230 nm, corresponding to cellulose (Via et al., 2003). This spectral profile expresses an opposition between the concentration of polyphenols and that of the non-structural carbohydrates (starch). For the spatial part, the magnitude bears the most weight. The dark areas of the image therefore correspond to heterogeneous areas in terms of carbohydrate concentration, while the light areas of the image correspond to more homogeneous areas in polyphenol concentration. It can then be deduced that the polyphenols present in the heartwood are distributed more evenly than the carbohydrates present in the sapwood. On the outer part of the heartwood, a light circle can be observed which may correspond to a defence mechanism of the living part of the wood (sapwood) against fungal colonization by producing these phenolic compounds (Niamké et al., 2011). Inside the heartwood, the discontinuous light circles constitute layers of wood protection.

Image score 6:

The image displays a distribution of light and dark areas across the entire surface of the disk, as well as a white line outside the heartwood and two dark spots at the top and bottom. The spectral loadings exhibit a negative peak at 1138 nm corresponding to lignin. A positive peak appears at 1200 nm corresponding to starch (Williams and Norris, 1987). Two other positive peaks at 1426 nm and 2224 nm correspond to cellulose (Schwanninger et al., 2011). There are also two peaks not identified in the literature, one negative at 1540 nm and the other positive at 1888 nm. The peak at 1678 nm corresponds to polyphenols. A negative peak at 1930 nm corresponds to the OH absorption band of water (Schwanninger et al., 2011), and the negative peak at 2260 nm corresponds to hemicellulose (Schwanninger et al., 2011). As for spatial loadings, magnitude is the most important variable followed by coherence, and both loadings are positive. This indicates that the darkest areas exhibit strong heterogeneity and local anisotropy in terms of concentration of compounds such as cellulose, starch, and polyphenols. Conversely, local concentrations of hemicellulose, lignin, and water appear to be more isotropic and homogeneous. The white line may correspond to a polyphenol-rich exchange zone.

Image score 7:

The image shows a dark area extending on the outer part of the heartwood, separating lighter areas located in the sapwood and centre of the heartwood. There are also some white spots present in both the heartwood and sapwood. The spectral loadings show a negative peak at 1132 nm corresponding to lignin (Schwanninger et al., 2011), as well as a small positive peak at 1216 nm, which corresponds to the second harmonic of $CH_2$ in cellulose (Schwanninger et al., 2011). Another positive peak at 1384 nm is related to the combination band of OH and CH (Schwanninger et al., 2011), while a negative peak at 1636 nm is associated with the first harmonic of OH in cellulose. A positive peak at 1678 nm is attributable to polyphenols, while an unidentified positive peak appears at 1886 nm. Finally, a negative peak at 1942 nm corresponds to the peak of water OH (Schwanninger et al., 2011), while two positive peaks appear at 2128 nm and 2266 nm, associated with lignin (Schwanninger et al., 2011). The spectral profile reveals contrasting concentrations of cellulose, lignin, and polyphenols. The spatial loadings indicate that magnitude is the variable with the strongest weight. Thus, the lighter areas are more heterogeneous in terms of cellulose, while the darker areas are more homogeneous in terms of lignin and polyphenols. In fact, this dark area corresponds to a zone of heartwood formation, where the sapwood cells transform into heartwood. This process results in a morphological evolution of cells, forming tyloses and tannin deposits. The white spots may correspond to partially perforated cell walls of vessel cells.

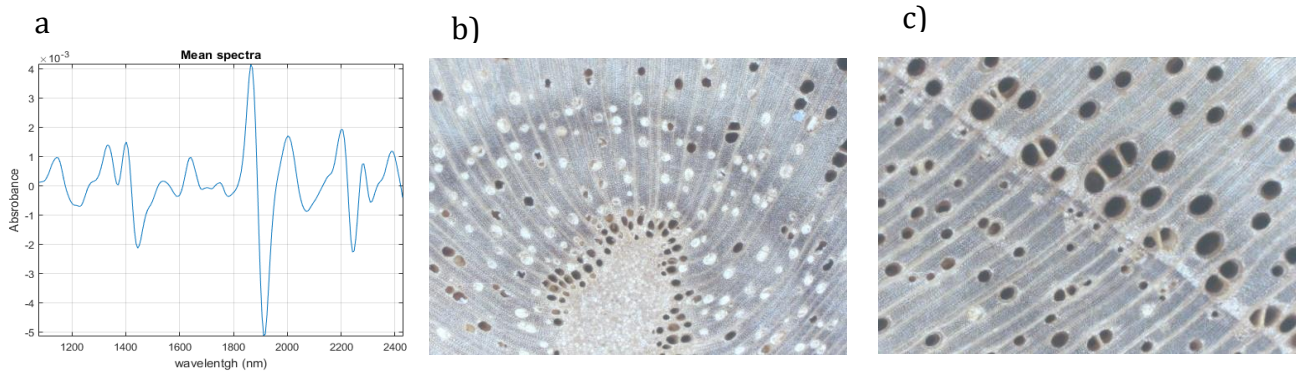

*Figure 3: a) average spectrum of all teak disk spectra b) microscopic image of the pith c) microscopic image of the area around an annual growth ring*

### 4.2 Second case study

#### 4.2.1 Principal component analysis

Three components were needed to explain 99% of the variance.

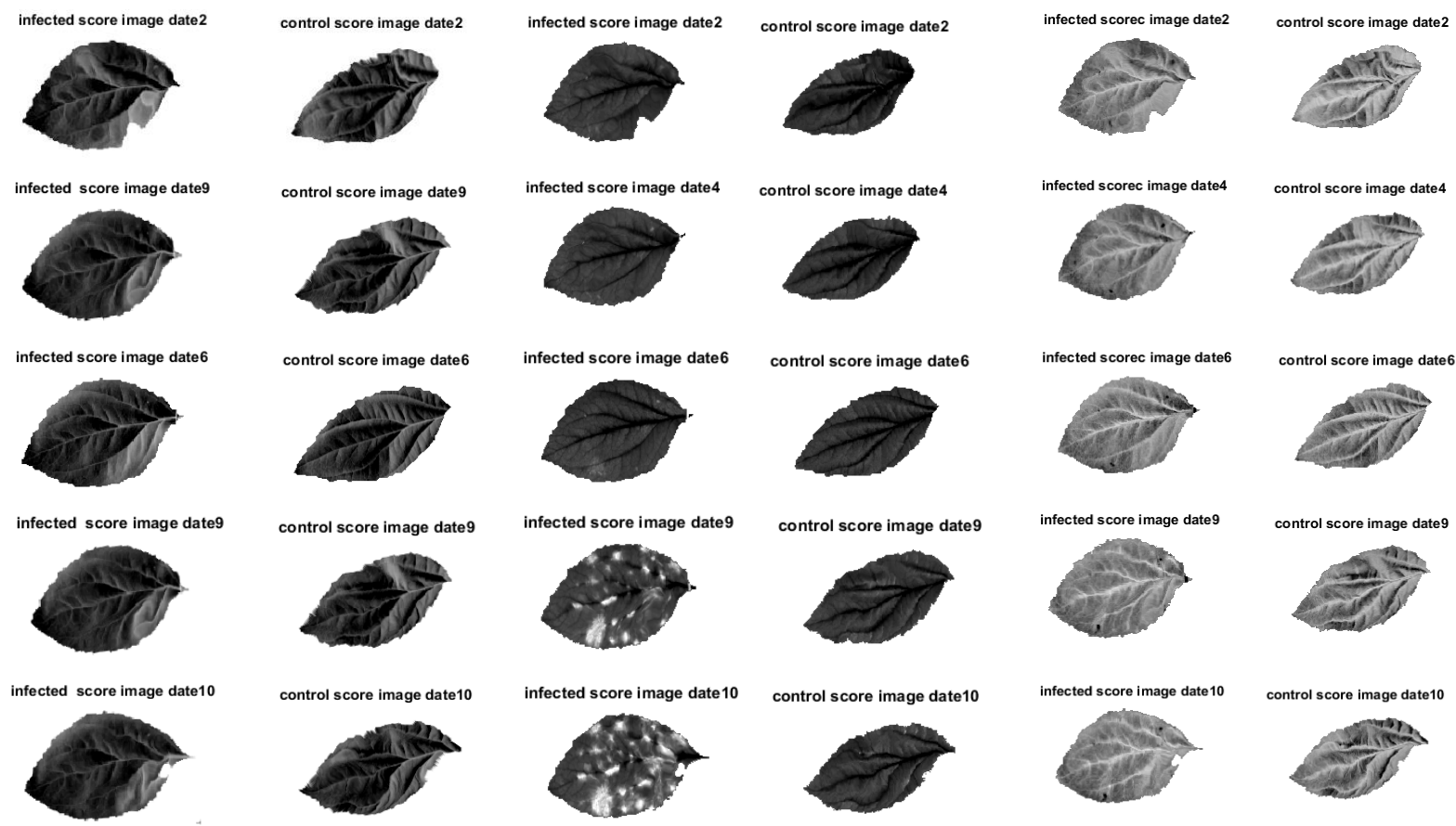

*Figure 4 : Projection of the calibration images onto the first 3 PCA loadings. Image scores (a), image scores (b) and image scores (c) represent the projection onto the first, second and third PCA loadings respectively.*

Figure 4 shows the score images resulting from the PCA performed on the 2080 spectra of the calibration set. These score images result from the projection of the hyperspectral images of the calibration set onto the first 3 loadings of the PCA. The score images of component 1, Figure

4(a), do not allow differentiation between infected and control leaves. Figure 5(a) represents the loadings of the PCA. The loadings of component 1 present values all the same sign and resemble the average spectrum of the leaves (not shown). This is typical of a global variation of the spectrum, due to a multiplicative effect. This is related to the variation in light scattering in the leaves. Therefore, scattering does not seem to be discriminative between control and inoculated leaves. This first observation indicates that the reflectance level alone is not sufficient, and that discrimination must use a more complex model. The loadings of component 2, Figure 5(b), show peaks at 1440 nm and 1940 nm and a slope between 1000 and 1400 nm. The peaks can be attributed to the absorptions of hydroxy OH groups. The slope can be attributed to a vegetation-specific scattering (Phelan et al., 2002). The corresponding scores images, Figure 4(b), clearly show the disease at 9 and 10 days post inoculation. We can therefore hypothesize that the development of the fungus *in planta* affects the state of the water and its content in the leaves (peaks at 1440 and 1940 nm) as well as the tissue structure (slope between 1000 and 1400 nm). This result is consistent with the study by Gorretta et al. (2019) conducted on the same data. These findings demonstrate the need to combine information from physical and chemical effects. The loadings of component 3, Figure 5(c), also present negative peaks around 1440 nm and 1940 nm. The corresponding score images, Figure 4(c), show a difference between the leaf blade and veins. These loadings appear to be carried by a difference in reflectance between these two tissues but do not allow to distinguish the images of the inoculated leaf from those of the control leaf.

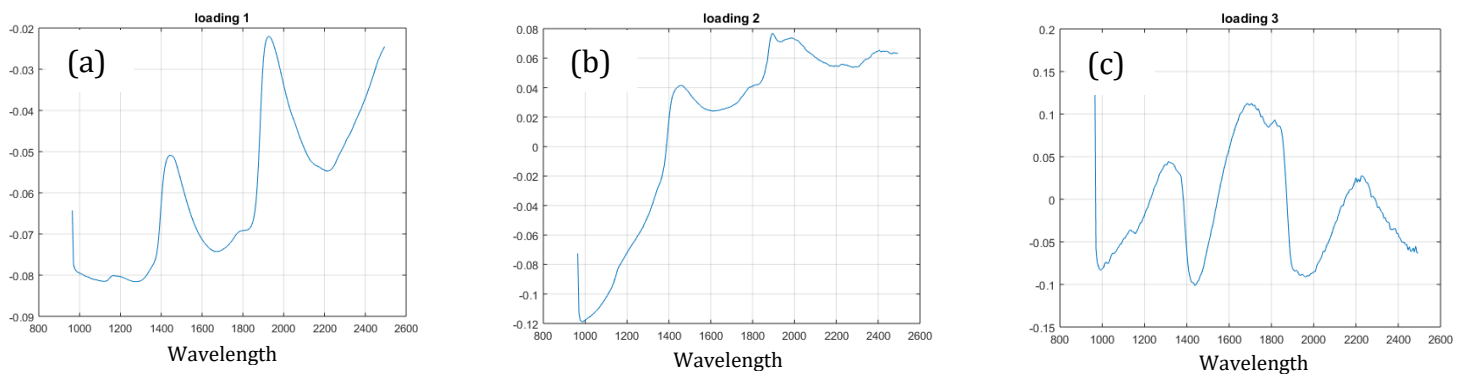

*Figure 5 : The loadings of the first three components PCA performed on 2080 spectra extracted from the images of the calibration set. a) loadings of the first component b) loadings of the second component c) loadings of the third component.*

## 4.2.2 –Spatial – spectral method parameterization

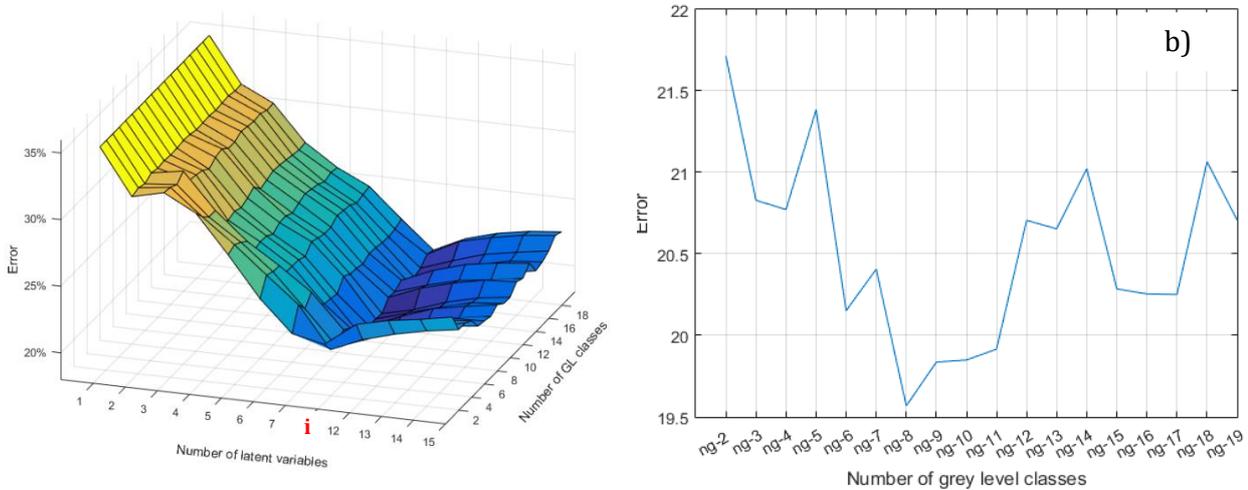

*Figure 6 : a) Evolution of the cross-validation error as a function of the number of latent variables and the number of grey level classes. (i) represents the average of the errors between 8 and 11 latent variables. b) The minimum cross-validation errors of the different models according to the number of grey level classes considered in the GLCM*

For the configuration of the spatial method, different numbers of classes of grey levels were considered (from 2 to 19). Figure 6(a) shows the evolution of the cross-validation error as a function of the number of latent variables of ROSA-PLS and the number of classes of grey levels. This figure shows that all curves reach a minimum error around or below 22% for a number of latent variables between 8 and 11 (noted "**i**" in the figure). A typical behaviour is observed: the error decreases up to the optimal number of dimensions and then increases. This shows that the models have learned meaningful information. Figure 6(b) represents the minimum errors of the different models according to the number of classes of grey levels considered in the GLCM. On this figure, the minimum is obtained with 8 classes of grey levels. Below this level of details, the texture variations between the two tissue types are not noticeable. Beyond this level, the co-occurrence matrices become sparse or "hollow". Indeed, if NG being the number of classes of grey levels, the co-occurrence matrices are of dimension NG × NG. However, the sub-images used to populate them are of size 3 × 3 = 9 pixels. As NG increases, the maximum number of co-occurrences quickly becomes much lower than the dimension of the matrix, which becomes sparse. The resulting Haralick indices become then less informative. Despite this finding, the difference between the errors of the different models is small. Therefore, it can be said that the number of classes of grey levels does not have a great influence on the models' performances, if chosen in a reasonable range. The number of classes of grey levels equal to 8 with 9 latent variables seems to be the right compromise for this data and has therefore been retained.

#### 4.2.2.1 Analysis of the overall model

A ROSA-PLS-LDA calibration was performed using the previously selected parameters: 3x3 image size, 3 retained PCA components, 8 grey-level classes, 1 for co-occurrence distance, sum of the matrices from the four directions, and three spectral signatures. Table 2 shows the blocks selected by the method for each latent variable.

*Table 1: Order and type of blocks selected by ROSA-PLS-LDA with 8 greyscale classes and 9 latent variables.*

| Number of the latent variable | 1 | 2 | 3 | 4 | 5 | 6 | 7 | 8 | 9 |
|---|---|---|---|---|---|---|---|---|---|
| Block number | 1st bloc spectral | 1st bloc spectral | 1st bloc spectral | 2nd bloc spatial | 1st bloc spectral | 1st bloc spectral | 1st bloc spectral | 1st bloc spectral | 1st bloc spectral |
| % of cross-validation error | 35 % | 32% | 31% | 28% | 27% | 25% | 22% | 20% | 19% |

Table 2 shows that the ROSA-PLS-LDA method first selects three times in a row the same spectral block. This spectral block corresponds to the spectral signature obtained with the first SVD component of each sub-image. At this stage, the error decreases to 31%. Then, the model selects a spatial block at the 4th latent variable. This block corresponds to texture features extracted from the score images of the second PCA component. With this spatial block, the error decreases again to 28%. Then, each iteration selects again the same spectral block from the first SVD loading until the model reaches its minimum error (19%).

This table shows that spectral information is predominant, as only one spatial component appears in fourth position and reduces the error by 4%. The selection of the second spatial block in the model, which is based on texture indices of score images "2" (see Figure 5(b)), confirms the conclusion of section 4.3.1 of the principal component analysis, which states that this component is related to the presence of the fungus. Finally, the model requires both types of spatial and spectral information to predict the class of images, demonstrating a complementarity between these two types of information.

#### 4.2.2.2 Performance of the overall model

The performance is evaluated on the test set. Table 3 shows the discrimination error between healthy and infected areas, using 9 latent variables from the ROSA-PLS-LDA model. The test model shows an average error of 10.9%. The model predicts the infected class with more accuracy than the healthy class, which is probably due to the higher variability of the healthy class in the calibration set. This may be due to the uniform destruction of the cell tissue caused by the scab, which makes the infected spots more homogeneous than the tissue of the healthy leaf.

*Table 2 : Confusion matrix of the test set*

|  |  | Real Classes |  |  |
|---|---|---|---|---|
|  |  | Infected | Control | Error |
| Predicted classes | Infected | 573 | 86 | 13.0% |
|  | Control | 51 | 538 | 8.6% |
| Error |  | 8.1% | 13.7% | **10.9%** |

### 4.2.2.3 Performance of the overall model detailed by date

Table 4 shows the classification error of the global model for each test date using 9 latent variables. The results from date 3 show more errors in the control class than in the infected class. The results from date 5 show more errors in the infected class than in the control class. For date 11, all errors are made in the control class. The variation in performance across dates is attributed to how the global model is influenced by dates 3 and 5, where the gap between inter- and intra-class variability is smaller. This reinforces the previously established conclusion that scab lesions are more homogeneous than healthy leaf tissue. This table also shows that the average prediction error decreases as we move from early to later dates (date 3 = 12.7%, date 5 = 11.0%, date 11 = 9.1%). These errors may be partly due to labelling errors as the lesion positions were defined at a later date (date 11).

*Table 3 : Confusion matrices of the test set for the different dates*

|  |  | Real classes |  |  |
|---|---|---|---|---|
|  |  | Date 3 |  |  |
|  |  | Infected | Control | Error |
| Predicted classes | Infected | 188 | 33 | 14.9% |
|  | Control | 20 | 175 | 10.2% |
| Error |  | 9.6% | 15.8% | **12.7%** |
|  |  | Date 5 |  |  |
| Predicted classes | Infected | 177 | 15 | 7.8% |
|  | Control | 31 | 193 | 13.8% |
| Error |  | 14.9% | 7.2% | **11.0%** |

|                   |           | Date 11 |       |       |
|-------------------|-----------|---------|-------|-------|
| Predicted classes | Infected  | 208     | 38    | 15.4% |
|                   | Control   | 0       | 170   | 0.0%  |
| Error             |           | 0.0%    | 18.2% | **9.1%** |

*4.2.2.4 : Interpretation of discriminant vectors :*

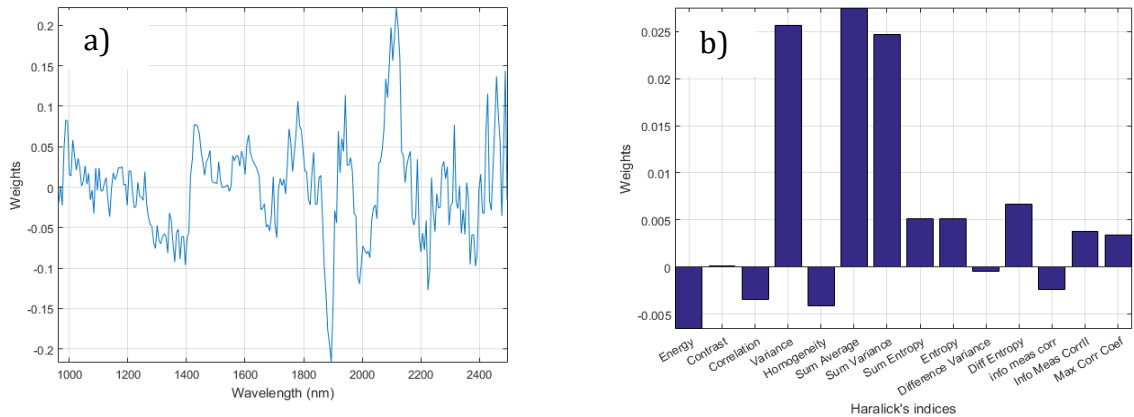

*Figure 7: Discriminant vectors. a) in the spectral domain (block 4). b) in the spatial domain (block 2).*

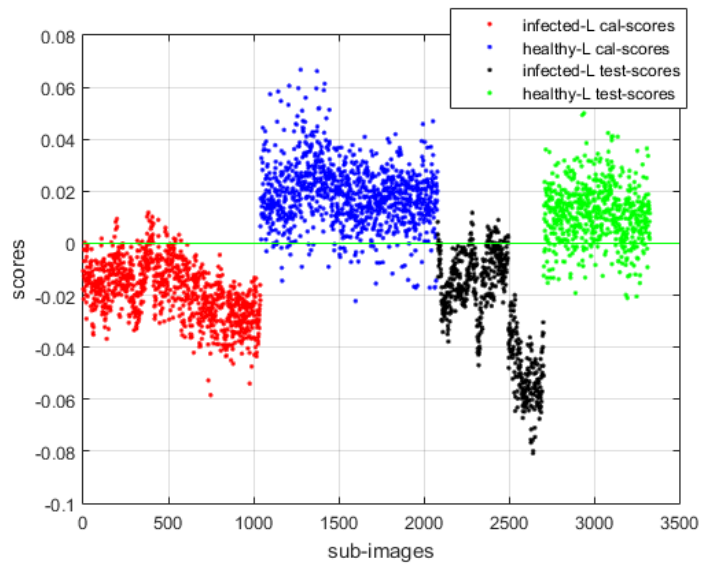

*Figure 8: ROSA-PLS-FDA scores: In red the calibration scores of the infected class, in blue the calibration scores of the control class, in black the test scores of the infected class and green the test scores of the control class.*

Figure 7 shows the discriminant vectors of the global model with 9 latent variables. Figure 8 shows the calibration and test set scores produced by this model. Figure 7 (a) represents the spectral part of the model. The spectral discriminant vector first shows a negative slope between 1100 nm and 1400 nm. Spectra that show a positive slope in this range produce negative scores and will therefore be classified as infected (see Figure 7). This type of spectral baseline is due to light scattering that may be related to leaf structure. Indeed, the mesophyll of a leaf is normally a layered structure in which light follows a linear behavior. When the fungus develops in a leaf, the cells lose their turgor (Gadoury and MacHardy, 1985), continuity, and sometimes the mycelium develops between the cells, causing a change in light scattering. The spectral discriminant vector also shows a succession of positive and negative peaks in the water absorption zone at 1940 nm. This type of spectral structure is related to a modification of the position and/or shape of the water absorption trough. This certainly reflects that the action of the fungus induces a change in the state of water. Finally, the spectral discriminant vector shows a positive peak at 2110 nm. This peak may be related to amides. This can be explained by the composition of the fungus, by the proteins (enzymes) secreted in the pathogenic mechanisms, or even by the defence mechanisms of the plant.

Figure 7 (b) describes the contribution of each of the Haralick indices in the classification. It appears that the most relevant characteristics are all related to a form of variance. These contributions are of positive sign. This means that they are related to the control class, which has positive scores (see Figure 8). This can be explained by the fact that control samples have structure, while apple scab spots have lost this structure due to the spread of the fungus in the tissues.

### 4.2.3 Model by date

A ROSA-PLS-LDA model was calculated and tested for each date. The parameters used were the same as for the global model. Table 5 represents the blocks selected by these models. It can be observed from this table that the models at early dates select spatial blocks first and predominantly. This can be explained by the fact that in the early phase, the main effect of the fungus is the modification of the optical properties of the leaves and therefore the texture observed on the reflectance images. This modification is certainly due to several physiological causes. The most important is certainly the release of intracellular water into intercellular spaces. This changes the optical continuity of the medium, and thus the diffusion of light (Jacquemoud and Ustin, 2019). However, the changes are more difficult to perceive on spectral tables because the chemical modifications are not yet at work at early stage of infection. Then, when the spots become visible, spectral information becomes predominant because there are many chemical phenomena that become perceptible (see the interpretation part of the discriminant vector of the global model). This result shows that the respective weights of the two types of information varies depending on the stage of the disease and that to build a predictive model for all dates, the calibration model must, on the one hand, contain the different stages of the pathology and on the other hand, use both types of information.

*Table 4 : Blocks selected by ROSA-PLS-FDA models calibrated on each date (yellow boxes are spatial blocks and green boxes are spectral blocks). The number in brackets indicates the number of the selected block.*

|         | 1        | 2        | 3        | 4        | 5        | 6        | 7        | 8        | 9        | 10       | 11       | 12       | 13       | 14       |
|---------|----------|----------|----------|----------|----------|----------|----------|----------|----------|----------|----------|----------|----------|----------|
| Date 2  | Spat(2)  | Spat(2)  | Spat(1)  | Spat(3)  | Spat(1)  | Spat(3)  | Spat(3)  | Spat(3)  | Spat(3)  | Spat(3)  | Spat(3)  | Spat(3)  | Spat(1)  | Spat(3)  |
| Date 3  | Spat(2)  | Spat(2)  | Spat(1)  | Spat(1)  | Spat(3)  | Spat(3)  | Spat(3)  | Spat(3)  | Spat(3)  | Spat(1)  | Spect(1) | Spect(1) | Spect(1) | Spect(1) |
| Date 4  | Spat(3)  | Spat(2)  | Spat(1)  | Spat(2)  | Spat(1)  | Spat(3)  | Spat(3)  | Spect(1) | Spect(1) | Spect(1) | Spect(1) | Spect(1) | Spect(1) | Spect(1) |
| Date 5  | Spat(2)  | Spat(2)  | Spat(3)  | Spat(1)  | Spat(3)  | Spat(1)  | Spect(1) | Spat(3)  | Spat(1)  | Spect(1) | Spect(1) | Spect(1) | Spect(1) | Spect(1) |
| Date 6  | Spat(2)  | Spat(2)  | Spat(3)  | Spat(1)  | Spat(1)  | Spat(3)  | Spat(3)  | Spat(3)  | Spat(1)  | Spat(2)  | Spat(3)  | Spect(1) | Spect(1) | Spect(1) |
| Date 9  | Spect(1) | Spect(1) | Spect(1) | Spect(1) | Spect(1) | Spect(1) | Spect(1) | Spect(1) | Spect(1) | Spect(1) | Spect(1) | Spect(1) | Spect(1) | Spect(1) |
| Date 10 | Spect(1) | Spect(1) | Spect(1) | Spect(1) | Spect(1) | Spect(1) | Spect(1) | Spect(1) | Spect(1) | Spect(1) | Spect(1) | Spect(1) | Spect(1) | Spect(1) |
| Date 11 | Spect(1) | Spect(1) | Spect(1) | Spect(1) | Spect(1) | Spect(1) | Spect(1) | Spect(1) | Spect(1) | Spect(1) | Spect(1) | Spect(1) | Spect(1) | Spect(1) |

## 4.3 General discussion on the methodology

This article describes a generic framework for combining the spatial and spectral information of a hyperspectral image. This process is flexible and can include different methods at each step of the scheme represented in Figure 1. In fact, in the first step, different methods can be used to extract the sub-images. They can be obtained with manual selection, with an automatic segmentation method (Borzov and Potaturkin, 2018), or by choosing a geometric shape centred on pixels of interest. In the second step, different approaches to dimensionality reduction are possible. The simplest is to select wavelengths based on domain knowledge or using a variable selection tool (Gauchi and Chagnon, 2001). Another approach is to summarize the spectrum to a set of statistics, such as the sum, mean, norm, or standard deviation, or distances, such as Euclidean distance, Mahalanobis distance, or a dedicated distance such as the Spectral Angle Mapper (Yang et al., 2008). The last approach is to use multivariate factor analysis methods such as principal component analysis, PCA (Wold et al., 1987), multivariate curve resolution alternating least squares, MCR-ALS (Tauler et al., 2002), or independent component analysis, ICA (Stone, 2002). These methods define a set of factors, a linear combination of wavelengths. The projection of spectra onto these factors reduces the spectrum to a set of scores. The third step, which is the extraction of spatial features, can rely on different image analysis techniques to produce indicators such as Linear Binary Pattern (Mu et al., 2008), Fourier coefficients, wavelet coefficients, Haralick indices, and structure tensors. Step 5 can also rely on different chemometric techniques that produce spectral loadings, such as PCA, MCR-ALS, or ICA, or statistical techniques that produce mean or standard deviation spectra. For step 6, different supervised and unsupervised data fusion methods can be used. Among unsupervised methods, SUM-PCA is a simple extension of PCA for the multi-block scenario (Smilde et al., 2003); MB-PCA or PCA consensus (Westerhuis et al., 1998) extracts global components and associated block contributions; CCSWA (Qannari et al., 2001) computes common and specific parts of the blocks. Supervised methods can be used when a response variable Y can be matched with the extracted sub-images to perform regression or discrimination; MB-PLS is an extension of PLS for multi-block scenarios (Wangen and Kowalski, 1989); the SO-PLS approach involves a series of PLS regression operations and matrix orthogonalization to sequentially extract information from different blocks (Næs et al., 2013); ROSA-PLS involves competing PLS components extracted in different blocks and selecting those most correlated with the response variable to be predicted (Liland et al., 2016). Each of these methods has advantages and disadvantages. The choice among these different methods depends on the images to be processed and the problem to be solved.

## 5 Conclusion

This article proposes a generic method for combining the spatial and spectral information of hyperspectral images. The case presented studies, illustrate the properties of this method, and demonstrate the relevance of the interpretations that can be derived from it. The first example uses a simple implementation with unsupervised data fusion using MBPCA. In this case study, the spatial part of the loadings changes mainly on the first four components, while on components 5 to 7, both the spatial and spectral loadings change. Components 5 to 7 revealed changes in the distribution of chemical compounds occurring within the wood disk. This shows that the same chemical information translates very differently in space, and this complementarity highlights phenomena that are invisible in purely spectral space.

The second case study, on apple scab data, presents a supervised application using ROSA-PLS-LDA. It enabled the prediction of the classes of infected and control image patches. The overall model containing different disease stages has an average performance of 11%. To achieve this performance, the proposed method used 8 spectral blocks and one spatial block. In this model, spectral information is predominant over spatial information. Nevertheless, in the date-based model, the importance of spatial and spectral information changes depending on the disease development stage. At the early and intermediate stages of the disease, spatial blocks were selected first. However, at the advanced stage of the disease, a single spectral block was selected.

The combined analysis of the spatial and spectral properties of a hyperspectral image provides an enhanced understanding of the phenomena being examined. The spectral part reveals chemical modifications, while the spatial part indicates how these modifications are distributed in space.

# 6 Références

Amigo, J.M., Babamoradi, H., Elcoroaristizabal, S., 2015. Hyperspectral image analysis. A tutorial. Analytica Chimica Acta 896, 34–51. https://doi.org/10.1016/j.aca.2015.09.030

Bigun, J., Bigun, T., Nilsson, K., 2004. Recognition by symmetry derivatives and the generalized structure tensor. IEEE Transactions on Pattern Analysis and Machine Intelligence 26, 1590–1605. https://doi.org/10.1109/TPAMI.2004.126

Borzov, S.M., Potaturkin, O.I., 2018. Spectral-Spatial Methods for Hyperspectral Image Classification. Review. Optoelectron.Instrument.Proc. 54, 582–599. https://doi.org/10.3103/S8756699018060079

Brown, M.T., Tinsley, H.E.A., 1983. Discriminant Analysis. Journal of Leisure Research 15, 290–310. https://doi.org/10.1080/00222216.1983.11969564

Doeswijk, T.G., Smilde, A.K., Hageman, J.A., Westerhuis, J.A., van Eeuwijk, F.A., 2011. On the increase of predictive performance with high-level data fusion. Analytica Chimica Acta, A selection of papers presented at the 12th International Conference on Chemometrics in Analytical Chemistry 705, 41–47. https://doi.org/10.1016/j.aca.2011.03.025

Feng, Y.-Z., Sun, D.-W., 2012. Application of Hyperspectral Imaging in Food Safety Inspection and Control: A Review. Critical Reviews in Food Science and Nutrition 52, 1039–1058. https://doi.org/10.1080/10408398.2011.651542

Filzmoser, P., Liebmann, B., Varmuza, K., 2009. Repeated double cross validation. Journal of Chemometrics 23, 160–171. https://doi.org/10.1002/cem.1225

Folch-Fortuny, A., Prats-Montalbán, J.M., Cubero, S., Blasco, J., Ferrer, A., 2016. VIS/NIR hyperspectral imaging and N-way PLS-DA models for detection of decay lesions in citrus fruits. Chemometrics and Intelligent Laboratory Systems 156, 241–248. https://doi.org/10.1016/j.chemolab.2016.05.005

Gadoury, D.M., MacHardy, W.E., 1985. Negative geotropism in Venturia inaequalis. Phytopathology 75, 856–859.

Gauchi, J.-P., Chagnon, P., 2001. Comparison of selection methods of explanatory variables in PLS regression with application to manufacturing process data. Chemometrics and Intelligent Laboratory Systems, PLS Methods 58, 171–193. https://doi.org/10.1016/S0169-7439(01)00158-7

Ghassemian Yazdi, M.H., 1988. On line object feature extraction for multispectral scene representation. Theses and Dissertations Available from ProQuest 1–71.

Gorretta, N., Nouri, M., Herrero, A., Gowen, A., Roger, J.-M., 2019. Early detection of the fungal disease "apple scab" using SWIR hyperspectral imaging, in: 2019 10th Workshop on Hyperspectral Imaging and Signal Processing: Evolution in Remote Sensing (WHISPERS). pp. 1–4. https://doi.org/10.1109/WHISPERS.2019.8921066

Griffiths, P.R., Dahm, D.J., 2007. Continuum and Discontinuum Theories of Diffuse Reflection, in: Handbook of Near-Infrared Analysis. CRC Press.

Haralick, R.M., 1979. Statistical and structural approaches to texture. Proceedings of the IEEE 67, 786–804. https://doi.org/10.1109/PROC.1979.11328

Imani, M., Ghassemian, H., 2020. An overview on spectral and spatial information fusion for hyperspectral image classification: Current trends and challenges. Information Fusion 59, 59–83. https://doi.org/10.1016/j.inffus.2020.01.007

Jacquemoud, S., Ustin, S., 2019. Leaf Optical Properties. Cambridge University Press.

Khan, M.J., Khan, H.S., Yousaf, A., Khurshid, K., Abbas, A., 2018. Modern Trends in Hyperspectral Image Analysis: A Review. IEEE Access 6, 14118–14129. https://doi.org/10.1109/ACCESS.2018.2812999

Klema, V., Laub, A., 1980. The singular value decomposition: Its computation and some applications. IEEE Transactions on Automatic Control 25, 164–176. https://doi.org/10.1109/TAC.1980.1102314

Kortüm, G., 2012. Reflectance Spectroscopy: Principles, Methods, Applications. Springer Science & Business Media.

Lelong, C.C.D., Pinet, P.C., Poilvé, H., 1998. Hyperspectral Imaging and Stress Mapping in Agriculture: A Case Study on Wheat in Beauce (France). Remote Sensing of Environment 66, 179–191. https://doi.org/10.1016/S0034-4257(98)00049-2

Liland, K.H., Næs, T., Indahl, U.G., 2016. ROSA—a fast extension of partial least squares regression for multiblock data analysis. Journal of Chemometrics 30, 651–662. https://doi.org/10.1002/cem.2824

M Hanafi, EM Qannari,, B Jaillais, n.d. Multi-block and three-way data analysis.

Medioni, G., 2000. A Computational Framework for Feature Extraction and Segmentation [WWW Document].

Mu, Y., Yan, S., Liu, Y., Huang, T., Zhou, B., 2008. Discriminative local binary patterns for human detection in personal album, in: 2008 IEEE Conference on Computer Vision and Pattern Recognition. pp. 1–8. https://doi.org/10.1109/CVPR.2008.4587800

Næs, T., Tomic, O., Afseth, N.K., Segtnan, V., Måge, I., 2013. Multi-block regression based on combinations of orthogonalisation, PLS-regression and canonical correlation analysis. Chemometrics and Intelligent Laboratory Systems 124, 32–42. https://doi.org/10.1016/j.chemolab.2013.03.006

Nouri, M., Gorretta, N., Vaysse, P., Giraud, M., Germain, C., Keresztes, B., Roger, J.-M., 2018. Near infrared hyperspectral dataset of healthy and infected apple tree leaves images for the early detection of apple scab disease. Data in Brief 16, 967–971. https://doi.org/10.1016/j.dib.2017.12.043

Qannari, E.M., Courcoux, P., Vigneau, E., 2001. Common components and specific weights analysis performed on preference data. Food Quality and Preference 12, 365–368. https://doi.org/10.1016/S0950-3293(01)00026-X

Savitzky, A., Golay, M.J.E., 2002. Smoothing and Differentiation of Data by Simplified Least Squares Procedures. [WWW Document]. ACS Publications. https://doi.org/10.1021/ac60214a047

Schwanninger, M., Rodrigues, J.C., Fackler, K., 2011. A Review of Band Assignments in near Infrared Spectra of Wood and Wood Components. J. Near Infrared Spectrosc., JNIRS 19, 287–308.

Smilde, A.K., Westerhuis, J.A., de Jong, S., 2003. A framework for sequential multiblock component methods. Journal of Chemometrics 17, 323–337. https://doi.org/10.1002/cem.811

Stone, J.V., 2002. Independent component analysis: an introduction. Trends in Cognitive Sciences 6, 59–64. https://doi.org/10.1016/S1364-6613(00)01813-1

Tauler, R., Kowalski, B., Fleming, S., 2002. Multivariate curve resolution applied to spectral data from multiple runs of an industrial process. ACS Publications. https://doi.org/10.1021/ac00063a019

Via, B.K., Shupe, T.F., Groom, L.H., Stine, M., So, C.-L., 2003. Multivariate Modelling of Density, Strength and Stiffness from near Infrared Spectra for Mature, Juvenile and Pith Wood of Longleaf Pine (Pinus Palustris). Journal of Near Infrared Spectroscopy 11, 365–378. https://doi.org/10.1255/jnirs.388

Vitale, R., Hugelier, S., Cevoli, D., Ruckebusch, C., 2020. A spatial constraint to model and extract texture components in Multivariate Curve Resolution of near-infrared hyperspectral images. Analytica Chimica Acta 1095, 30–37. https://doi.org/10.1016/j.aca.2019.10.028

Wangen, L.E., Kowalski, B.R., 1989. A multiblock partial least squares algorithm for investigating complex chemical systems. Journal of Chemometrics 3, 3–20. https://doi.org/10.1002/cem.1180030104

Westerhuis, J.A., Kourti, T., MacGregor, J.F., 1998. Analysis of multiblock and hierarchical PCA and PLS models. Journal of Chemometrics 12, 301–321. https://doi.org/10.1002/(SICI)1099-128X(199809/10)12:5<301::AID-CEM515>3.0.CO;2-S

Williams, P., Norris, K., 1987. Near-infrared technology in the agricultural and food industries. Near-infrared technology in the agricultural and food industries.

Wold, S., Esbensen, K., Geladi, P., 1987. Principal component analysis. Chemometrics and Intelligent Laboratory Systems, Proceedings of the Multivariate Statistical Workshop for Geologists and Geochemists 2, 37–52. https://doi.org/10.1016/0169-7439(87)80084-9

Yang, C., Everitt, J.H., Bradford, J.M., 2008. Yield Estimation from Hyperspectral Imagery Using Spectral Angle Mapper (SAM).